# Lumina: An AI-Augmented Multiscale Material Informatics Framework for Extreme Aero-Chemo-Thermo-Mechanical Regimes

Pradeep Kumar Seshadri[#*], Vigneshwaran N[$], Sudaroli Dhananjeyan[!], S. Karthikeyan[^], Navbila K[$], Sridhar S[$], Subhadevi K[$], Hari Sree Charan H[^], A Abdul Azeez[^], Jeswin Mickle[!], Harsha C[!]

[#] Department of Aerospace Engineering, IIT Madras, Chennai – 600036, India

[$]Artificial Intelligence and Data Science, Kongu Engineering College, Perundurai, Erode, Tamil Nadu, India

[^]Department of Aerospace Engineering, B.S. Abdur Rahman Crescent Institute of Science & Technology, Chennai – 600048, India

[!] Computer Science and Engineering Department, T John Institute of Technology, Bangalore – 560068, India

*Corresponding Author's E-mail: pkseshadri@smail.iitm.ac.in

## ABSTRACT

**Predictive simulations and experimental design involving extreme aero-chemo-thermo-mechanical regimes require high-fidelity material representation across diverse physical states. However, data for metals, polymers, and propellants, explosives, and pyrotechnics (PEP) remain fragmented, obstructing traceability for formulators, experimentalists, and simulation engineers. This work introduces Lumina, a modular Python-based informatics framework that centralizes multiscale material data—from atomistic simulation datasets to macro-scale experimental records—within a unified repository. Lumina employs a hierarchical XML-based schema and a dynamic runtime parsing mechanism to enable schema-independent parameter extraction. Beyond storage, the platform provides computational modules to visualize model fits, allowing experimentalists to optimize design of experiments (DoE) and formulators to validate chemical behaviors against benchmarks. This structured architecture serves as a high-fidelity pipeline for training machine learning models and enhancing the accuracy of predictive simulations. To streamline multi-disciplinary workflows, Lumina integrates a conversational AI assistant for intelligent material retrieval and natural language querying. By consolidating multiscale data into an extensible ecosystem, Lumina provides a scalable foundation for data-driven discovery and predictive modeling in advanced defense and aerospace engineering.**

## NOMENCLATURE

DB – Runtime material database; XML – Extensible Markup Language; API – Application Programming Interface ; $\rho$ = Density; $U_s$ = Shock Velocity; $U_p$ = Particle Velocity; P = Pressure; V = Volume; $V/V_o$ = Volume Ratio

# 1. INTRODUCTION

The development of materials for extreme aero-chemo-thermo-mechanical regimes requires a seamless continuum between chemical formulation, experimental characterization, and high-fidelity predictive simulation. Whether designing structural alloys or complex propellants, explosives, and pyrotechnics (PEP), the ability to accurately model material response depends on the accessibility of multiscale data—from atomistic benchmarks to macro-scale shock relationships. However, a significant bottleneck persists because these critical datasets are traditionally fragmented across disparate literature, proprietary reports, and static simulation input files. This fragmentation creates a disconnect between the formulators who develop new energetic compositions, the experimentalists who characterize their dynamic response, and the simulation engineers who deploy this data in numerical solvers. Without a unified informatics framework, traceability is lost, and the "rigid" nature of hard-coded material definitions makes the iterative Design of Experiments (DoE) and model calibration process inefficient and prone to error. To address these multidisciplinary challenges, this work introduces *Lumina*: a modular Python-based informatics ecosystem designed to centralize and activate material data for extreme environment research. *Lumina* transitions material representation from static text-based entries to a dynamic, activate-on-demand architecture. By utilizing a hierarchical XML-based schema and a runtime database builder, the framework provides a transparent pipeline where chemical properties, combustion parameters, and mechanical Equations of State (EOS) are unified. This architecture not only enhances the accuracy of predictive simulations but also empowers formulators and designers with automated visualization and AI-assisted retrieval tools, providing a scalable foundation for data-driven discovery in defense and aerospace engineering.

# 2. RELATED WORKS

Traditionally, material response in extreme regimes is modeled using empirical macro-scale relationships supported by thermochemical codes like NASA CEA or EXPLO5 [1] and modern MATLAB-based tools like HEMSim [2] to bridge theoretical gaps. However, high-fidelity design increasingly requires integrating multiscale data from atomistic studies and mesoscale dynamics—such as energy localization—to inform continuum behavior [3], [4], [5], [6], [7] yet these datasets remain fragmented for experimentalists and modelers. While traditional frameworks like OpenFOAM [8] and LS-DYNA [9] offer powerful solvers, their complex, non-modular architecture hinders seamless data reuse and the integration of diverse material formats. To address these bottlenecks, we introduce Lumina, an AI-augmented informatics ecosystem and the first simulation-ready database designed to couple chemical formulation with mechanical solvers through a lightweight, "activate-on-demand" architecture. Building on the modular, header-only principles of our *cpp-material-engine*, Lumina utilizes a hierarchical XML-based schema and a conversational AI assistant [10], [11] to centralize disparate multiscale records. This unified framework transforms fragmented data into a scalable pipeline [12] for predictive modeling and discovery across the full spectrum of aero-chemo-thermo-mechanical regimes.

# 3. MATERIAL AND METHODS

The development of Lumina is predicated on the transition from static, uncoupled data entries to an integrated, activate-on-demand informatics ecosystem. The methodology focuses on three primary pillars: a modular system architecture, a multi-layered hierarchical data structure, and a validation-centric computational layer.

## 3.1 Overall System Architecture

*Lumina* is designed as a modular and decoupled framework that separates data storage from runtime processing. This ensures flexibility, allowing analytical and visualization components to function independently of the storage format. The system integrates key engines, including an XML parser, runtime database builder, parameter extraction module, and model management unit for efficient material data processing.

### *3.1.1 System Architecture Overview*

The proposed system adopts a modular and scalable architecture for efficient management of material property database associated with extreme environment conditions. It integrates XML-based data structuring, a PostgreSQL database, dynamic visualization, a PyQt6-based GUI, and an AI-powered chatbot. Experimental data is transformed into hierarchical XML, validated, and stored in the database, enabling structured access and retrieval. The GUI supports material exploration, visualization, comparison, and high-resolution plotting, along with features such as reference tracking and simulation-ready XML export. An AI chatbot using a locally deployed LLM (Ollama with Llama 2) enables natural language queries [13] for efficient data access. This integrated pipeline ensures seamless data flow from ingestion to analysis. The overall workflow is illustrated in Fig. 1.

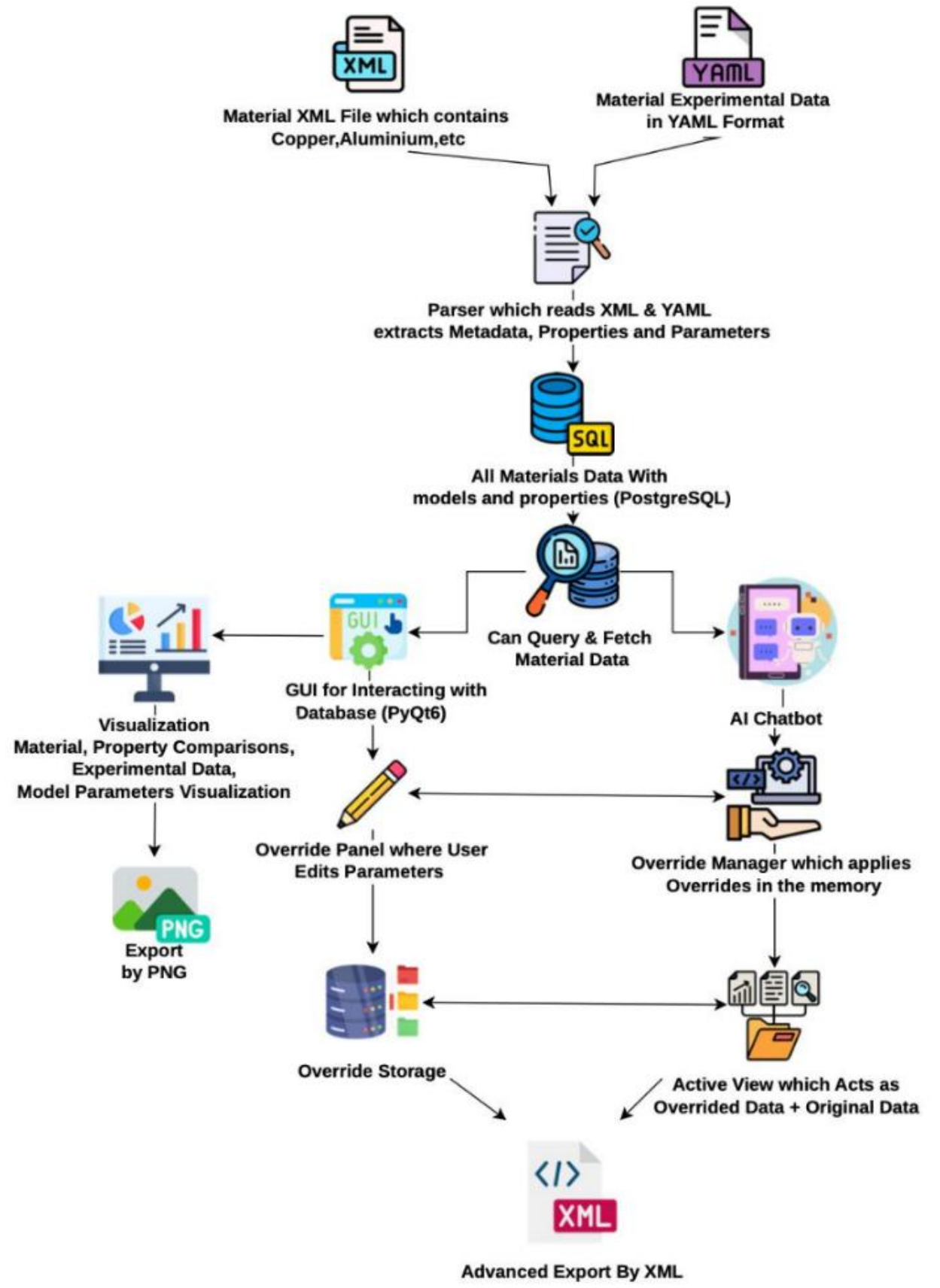

Fig 1: Material Data Processing and AI-Based Query Workflow

*3.1.2 Computational Engine Architecture*

The cpp-material-engine is a lightweight, modular backend for physics-based material evaluation using XML inputs. It is a header-only, stateless C++ engine with a custom parser, ensuring deterministic, dependency-free, and thread-safe execution. The design enables scalability, extensibility, and efficient integration with HPC systems Fig. 2.

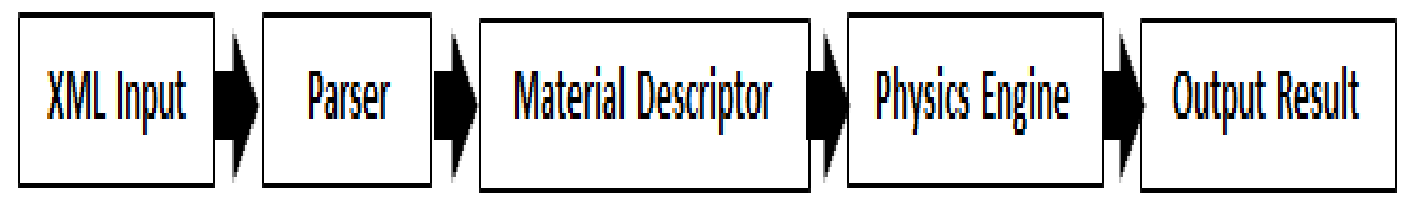

Fig. 2: Workflow of cpp-material-engine from XML input to computed output

*3.1.3 Material Model Visualization and Analysis Framework*

The Python-based visualization tool integrates multiple material models with a YAML-based material database. It features a modular design with components for data processing, validation, GUI, and material model evaluation. The system supports dynamic visualization and comparison of theoretical and experimental data, enabling efficient analysis of material behavior under extreme conditions.

### 3.2 Hierarchical XML-Based Material Data Structure

*Lumina* uses a hierarchical XML schema with eleven categories to represent material behavior. The database contains high-quality data from literature and experiments. The categories include:

1. **Structure and Formulation** (Chemical composition and multiscale descriptors)
2. **Physical, Chemical, and Thermal Properties**
3. **Detonation and Sensitivity Characteristics** (Critical for PEP applications)
4. **Mechanical, Electrical, and Toxicity Profiles**
5. **Models** (EOS, Strength, Reaction kinetics, etc.)

This structured categorization ensures consistent parameter naming, unit specification, and symbolic representation. *Lumina* ensures consistency in parameters, units, and symbols, while maintaining traceability of data sources from atomistic to macro-scale level data.

### 3.3 YAML-Based Experimental Data Structure

The system uses a YAML format to store experimental data in a simple and readable way, complementing XML-based classification. The structure includes:

1. **Metadata** (Material information, experiment type, model type, and references)
2. **Dataset Information** (Version details and creation date for reproducibility)
3. **Units** (Standardized units for all experimental parameters)
4. **Experimental Data Points** (Measured values such as shock velocity, particle velocity,

pressure, and density, along with labels and symbols for visualization)

This representation ensures clarity, consistency, and easy interpretation of experimental data, while supporting visualization and efficient integration with the system pipeline.

## 3.4 Material Descriptor and Data Structures

### *3.4.1 Internal Representation*

The parsed XML data is converted into lightweight C++ structures known as material descriptors. These structures store parameters for different models in a structured and accessible format. The design emphasizes simplicity and efficiency, enabling fast access during computations while maintaining readability and maintainability.

### *3.4.2 Memory Layout Considerations*

Efficient memory layout is critical for HPC performance. The engine uses compact data structures to minimize memory usage and improve cache locality. By avoiding unnecessary allocations and using contiguous storage where possible, the engine ensures high performance during repeated evaluations in simulation loops.

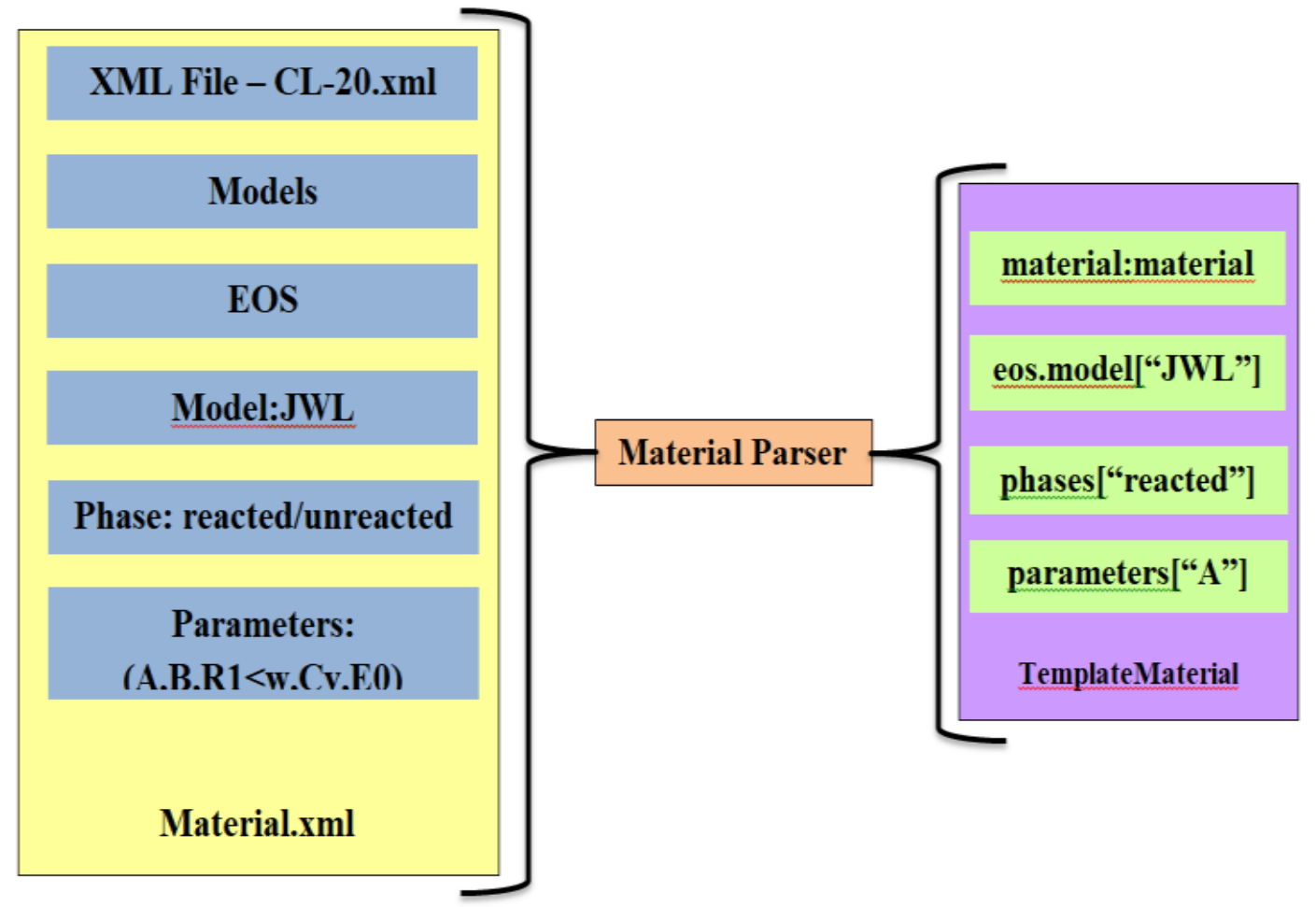

Fig 3: XML-C++ Mapping

### *3.4.3 Lightweight Object Design*

The cpp-material-engine ensures that objects are minimal and contain only essential data. This reduces overhead and improves performance. Lightweight objects are also designed to simplify data transfer between components and enhance compatibility with parallel computing environments.

### *3.4.4 Mapping XML to C++ Structures*

The mapping process converts XML nodes into corresponding C++ structures Fig 3. Each XML element is translated into a structured representation, preserving hierarchy and relationships. This mapping enables efficient access to parameters during evaluation and ensures consistency between input data and internal representation.

## 3.5 Performance and HPC Considerations

### *3.5.1 Thread Safety*

The stateless design ensures that multiple evaluations can be performed concurrently without conflicts. This makes the engine suitable for multi-threaded and distributed computing environments.

### *3.5.2 Memory Efficiency*

Memory usage is minimized through efficient data structures and avoidance of unnecessary allocations. This improves performance and enables the engine to handle large-scale simulations.

### *3.5.3 GPU/Parallel Compatibility*

The engine is designed to be compatible with GPU and parallel computing frameworks. Its stateless nature allows it to be easily integrated into parallel execution pipelines controlled by external systems.

### *3.5.4 Determinism and Reproducibility*

Deterministic behavior ensures consistent results across runs, which is critical for scientific validation. The engine avoids randomness and maintains strict control over computations.

## 3.6 Data State Management: Original, Override, and Active Views

To support the **Design of Experiments (DoE)** and iterative calibration, *Lumina* implements a unique three-layer data management mechanism. The **Original** layer serves as the immutable baseline derived from validated literature sources. Table 1 The **Override** layer allows formulators and experimentalists to input modifications—such as varying chemical concentrations or adjusting model parameters—to observe sensitivity. These are resolved into the **Active View.** Dynamic dataset for computation and visualization, enabling sandbox experimentation without affecting validated data.

Table 1. Data State Management

| Layer | Description |
|---|---|
| Original | Baseline validated data |
| Override | User-modified experimental values |
| Active | Overriden data for analysis |

## 4. INFORMATICS, VISUALIZATION AND AI-INTEGRATION

The framework integrates visualization, relational indexing, and an AI assistant to enable efficient analysis and intelligent material retrieval. This combination of multiscale data and machine learning capabilities makes *Lumina* a comprehensive solution to existing gaps.

### 4.1 Visualization and Analytical Interface

The system's visualization and analysis are enabled through a PyQt6-based GUI that provides an intuitive, tab-based interface for seamless interaction without requiring programming or database expertise. It integrates a centralized material browser, categorized property viewer with units and references, and a Matplotlib-based visualization module for analyzing temperature-dependent, pressure-dependent, and shock-related material behaviors using line, bar, and scatter plots. The system also supports multi-material comparison through graphical representations with high-resolution export, along with additional features such as literature reference viewing, data override and versioning, and simulation pre-processing.

### 4.2 AI-Integrated Query System

The system includes an AI Integrated Query Module to enable easy and intelligent interaction with the database. It allows users to perform queries using natural language instead of SQL. The module is powered by a locally deployed Llama 2 model integrated [14] through the Ollama framework, ensuring the system works completely offline. It acts as an intermediary by analyzing user queries, identifying intent, and extracting key entities such as materials, properties, and models. These inputs are then converted into optimized SQL queries, and the results are presented in a structured, human-readable format, making it easier for users to understand complex data without SQL expertise AI Work flow shown in Fig 4.

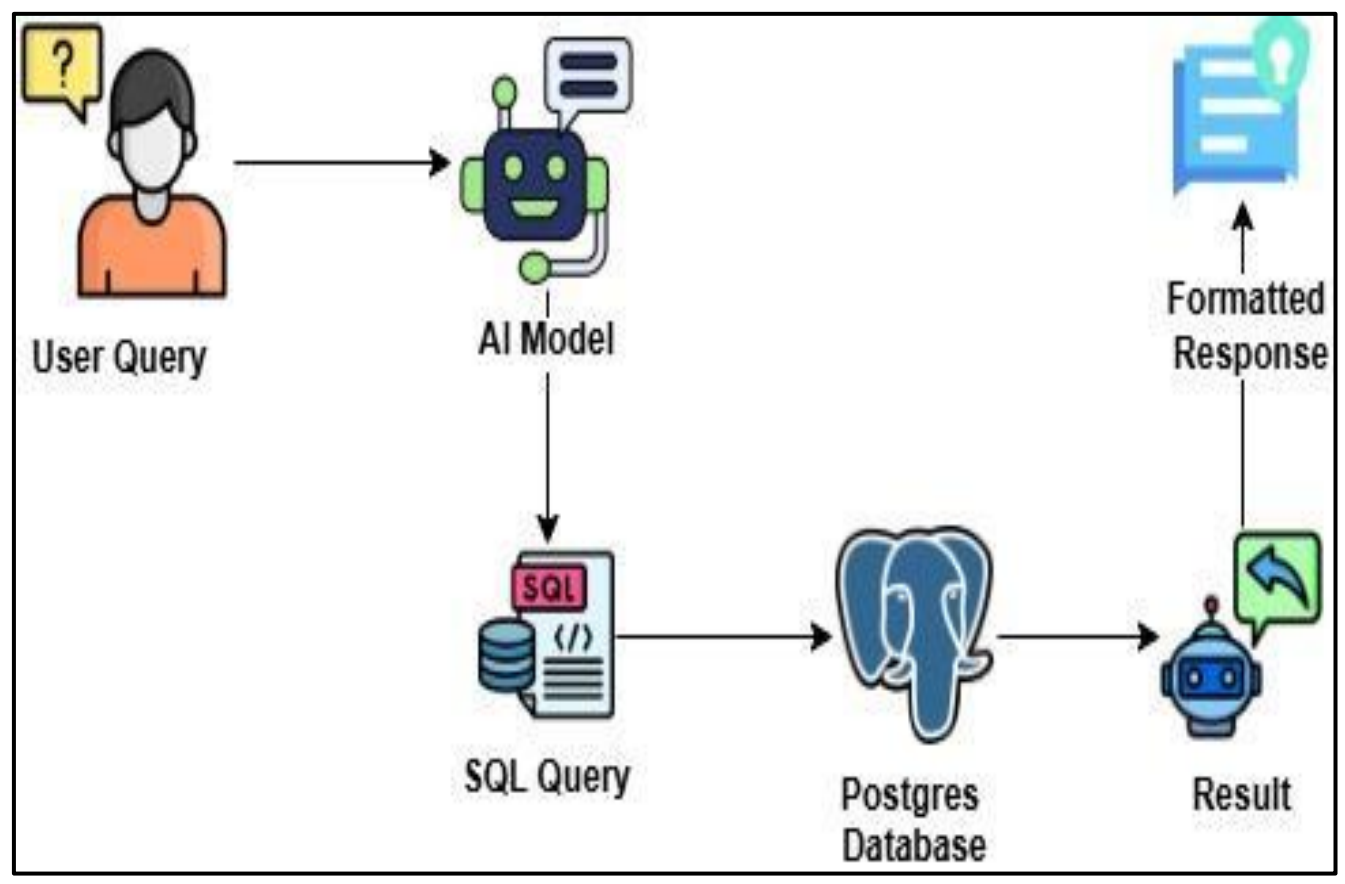

Fig 4: AI-Based Natural Language Query Processing

### 4.3 Model Management Framework

While our framework is designed to manage a wide array of material models—including plasticity, strength, and damage—we use the Equation of State (EOS) here to demonstrate how the system handles complex data relationships and automated calculations. The framework transforms these models from static entries into dynamic objects that link material parameters, theoretical physics, and experimental validation. To support this, we have digitized the Shock Hugoniot database from Los Alamos (LANL) reports and other open literature, storing raw experimental datasets alongside their corresponding empirical and theoretical fit information.

#### *4.3.1 Us–Up Relationship & Hugoniot Mapping*

The system manages the linear Shock Velocity ($U_s$) and Particle Velocity ($U_p$) relationship to characterize material behavior under shock. By encoding the underlying physics directly into the framework, the system can automate the transition from raw parameters to visual data.

1. **Integrated Parameter Storage:** The database stores material-specific constants—such as initial density ($\rho_0$), bulk sound speed ($C_0$), and the slope ($s$)—as interconnected attributes.

**2. Automated Calculation:** The system executes the conservation and state equations to derive pressure (P) and volume ($V/V_0$) changes:

$$P = \rho_o * U_p * U_s \quad (1)$$

$$U_s = C_o + s * U_p \quad (2)$$

**3. Experimental Benchmarking:** A core strength of the framework is its ability to compare theoretical curves against stored experimental data. This allows the user to immediately visualize how well the model fits real-world benchmarks, such as $P - V/V_0$ plots generated at desired intervals.

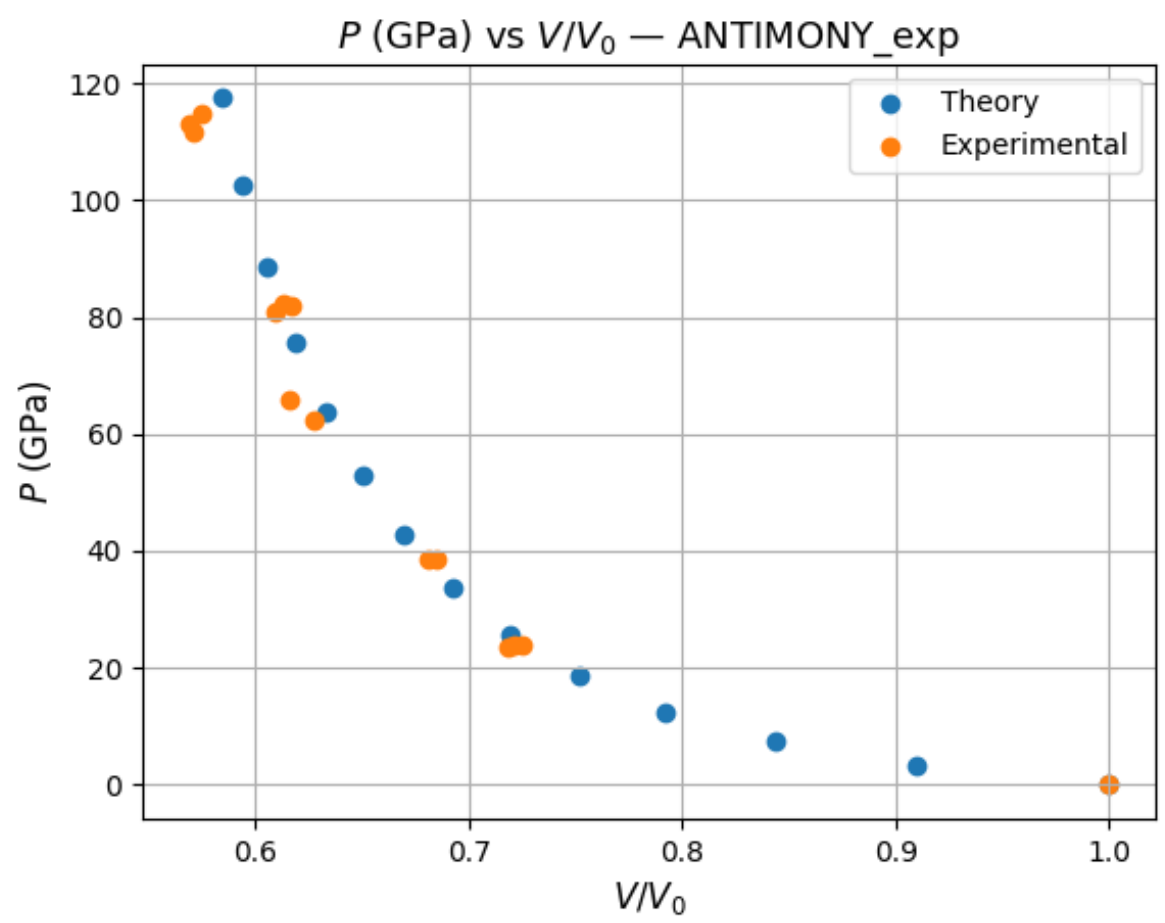

Fig 5: P-V plot from Us-Up EOS

*4.3.2 Multi-Scale Integration: Mie-Gruneisen EOS*

In extreme pressure and temperature regimes, experimental data is often inaccessible or difficult to obtain. For instance, consider the example of energetic materials like HMX or RDX. Our framework incorporates results from Molecular Dynamics (MD) studies to define the material's foundational physics. Because MD models are constantly evolving, our system is designed to ingest new reference curves as they are refined.

1. **Interactive Visualization of Equilibrium States:** The framework generates isothermal and isentropic projections on the $P - V/V_0$ plane. These reference states ($P_c$, $E_c$) are derived directly from MD-informed cold curves and lattice vibration data, allowing users to visualize the material's thermodynamic landscape before any experimental data is even applied.

$$P = P_c + \rho\Gamma(E - E_c) \quad (3)$$

2. **Complete EOS representation:** No EOS is complete without an accurate calculation of temperature (T). For materials like HMX, we have implemented the Menikoff-Sewell model, which partitions internal energy into a reference state and a nonlinear thermal component. To solve for T consistently with the Debye model and specific heat ($C_v$), the system employs a secant iteration method:

$$\Theta = 188.783\left(\frac{\rho}{\rho_0}\right)^{\Gamma_0} \exp\left[-\Gamma_1\left(\frac{\rho_0}{\rho} - 1\right) - \frac{1}{2}\Gamma_2\left(\frac{\rho_0^2}{\rho^2} - 1\right)\right] \quad (4)$$

**3. Multi-Model Database Management:** A core strength of the framework is its ability to manage diverse representations of the same property. For example, the Gruneisen parameter (Γ) can be stored as an MD-derived polynomial fit to capture anharmonic effects, or as a constant value for traditional models. This modularity allows researchers to instantly swap and compare different theoretical formulations to see their impact on the equilibrium state.

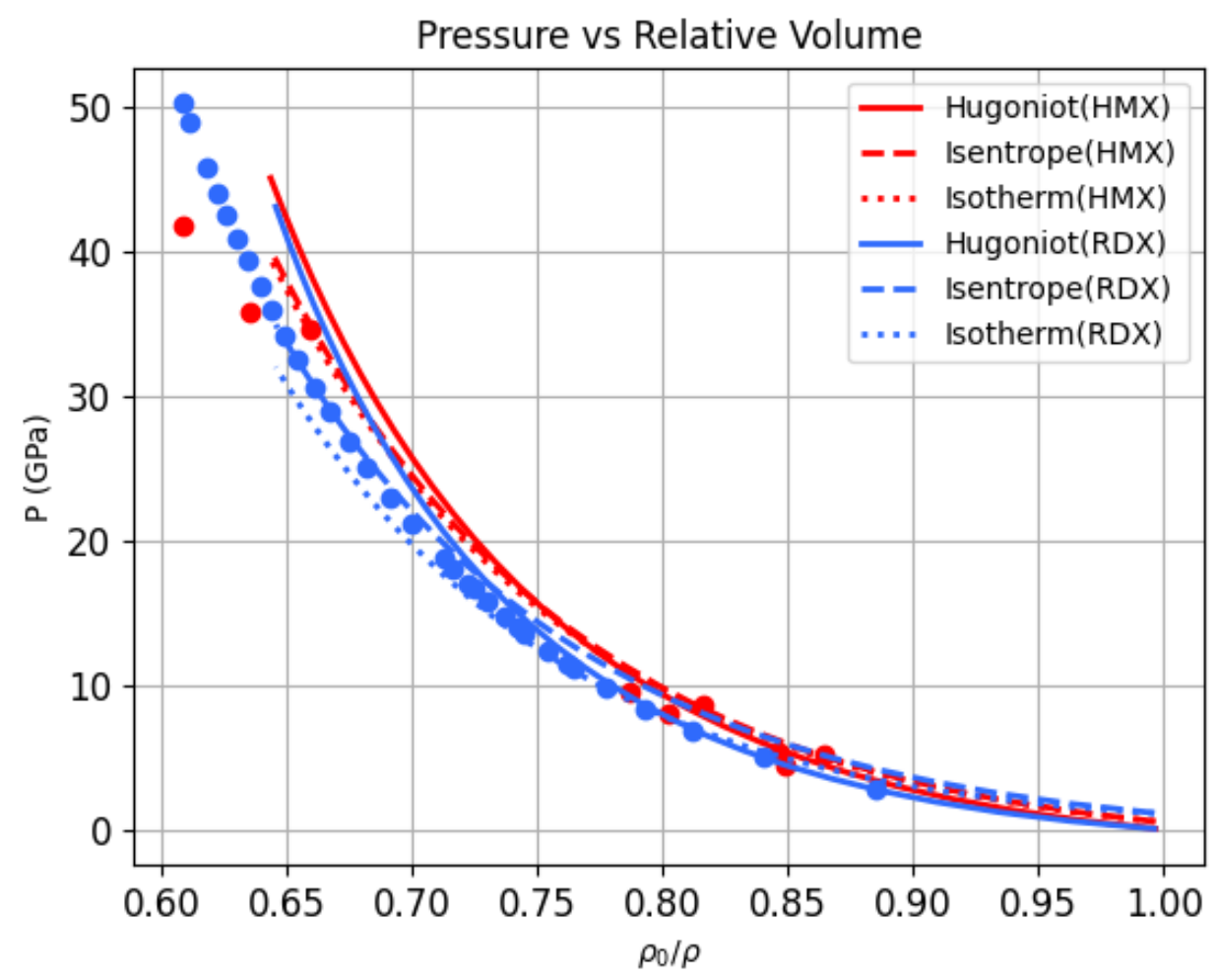

Fig 6: Mie-Gruneison Equation of state HMX

*4.3.3 JWL Equation of State for High Explosives*

While the Mie-Gruneisen framework handles the solid-state compression, the system also incorporates the JWL (Jones–Wilkins–Lee) Equation of State to model the detonation products and their subsequent expansion. Since JWL is an empirical, calibrated model, the framework highlights our database's ability to manage specific expansion regimes.

**1. Calibration-Ready Parameter Management:** The JWL model relies on material-specific constants (A, B, C, $R_1$, $R_2$, ω). In our framework, these are not just static values but part of a calibration-ready library. This allows users to fine-tune the parameters to match specific cylinder expansion (CYLEX) experimental data stored within the same database.

$$P(V) = A\,e^{\{-R_1 V\}} + B\,e^{\{-R_2 V\}} + C\,V^{\{-(1+\omega)\}} \tag{5}$$

2. **Expansion Regime Visualization:** A key feature of the framework is the ability to visualize the isentropic expansion path of detonation products. By solving the relationship between the relative volume (V) and internal energy (E), the system generates $P - V/V_0$ curves that transition from high-pressure detonation states to low-pressure expansion. This allows for a direct visual assessment of the work done by the explosive.

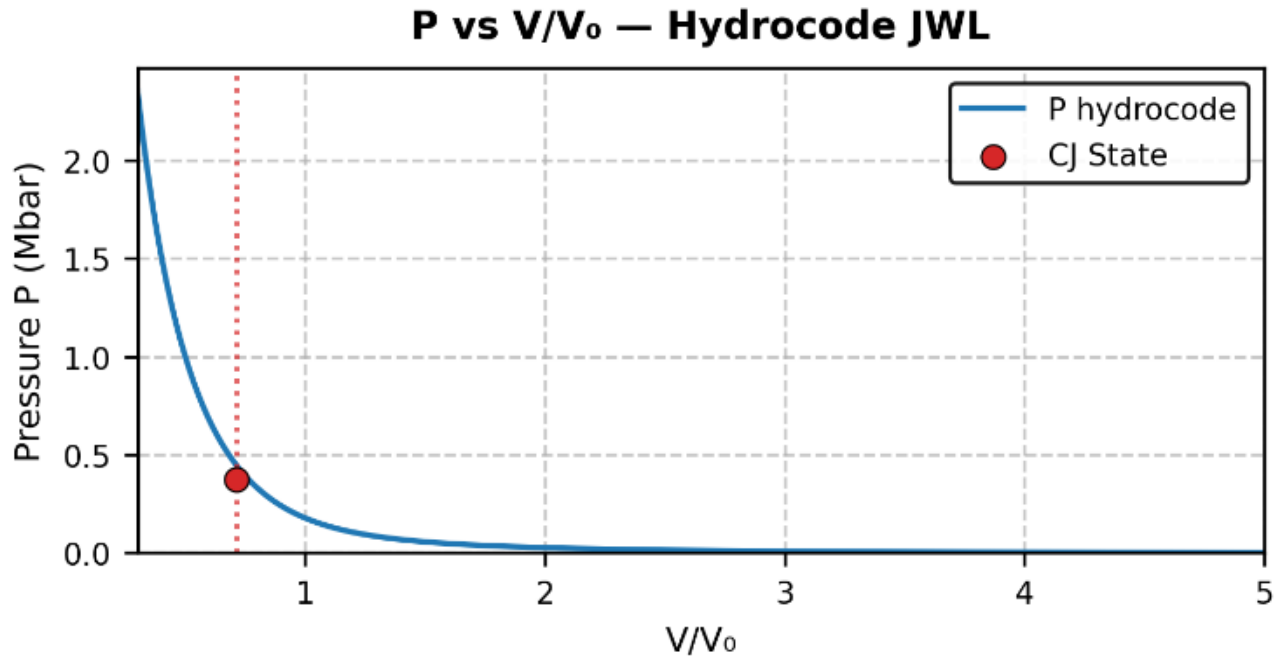

Fig 7: JWL Equation of State

**3. Multi-State Comparison Engine:** The framework's most powerful feature is the ability to overlay different EOS models. For a single high explosive like HMX, a user can visualize the Mie-Gruneisen "unreacted" solid state and the JWL "reacted" gas state on the same $P - V/V_0$ plot. This helps in understanding the energy release and the thermodynamic jump across the detonation front, demonstrating the framework's capacity to handle multi-phase material transitions.

## 5. RESULTS AND DISCUSSION

### 5.1 Coupling *Lumina* to Scientific Solvers

The proposed material engine is designed to operate independently of the underlying physics solver, enabling seamless integration across different simulation environments. This decoupled architecture allows material models to be defined externally through XML and dynamically loaded at runtime without modifying or recompiling the solver.

### 5.2 Runtime Parameter Extraction

The runtime database builder extracts model parameters directly from the XML material file and constructs the corresponding model representations during initialization. For example, the JWL equation of state is identified from the XML structure, and its parameters are mapped into internal data structures Fig 8.

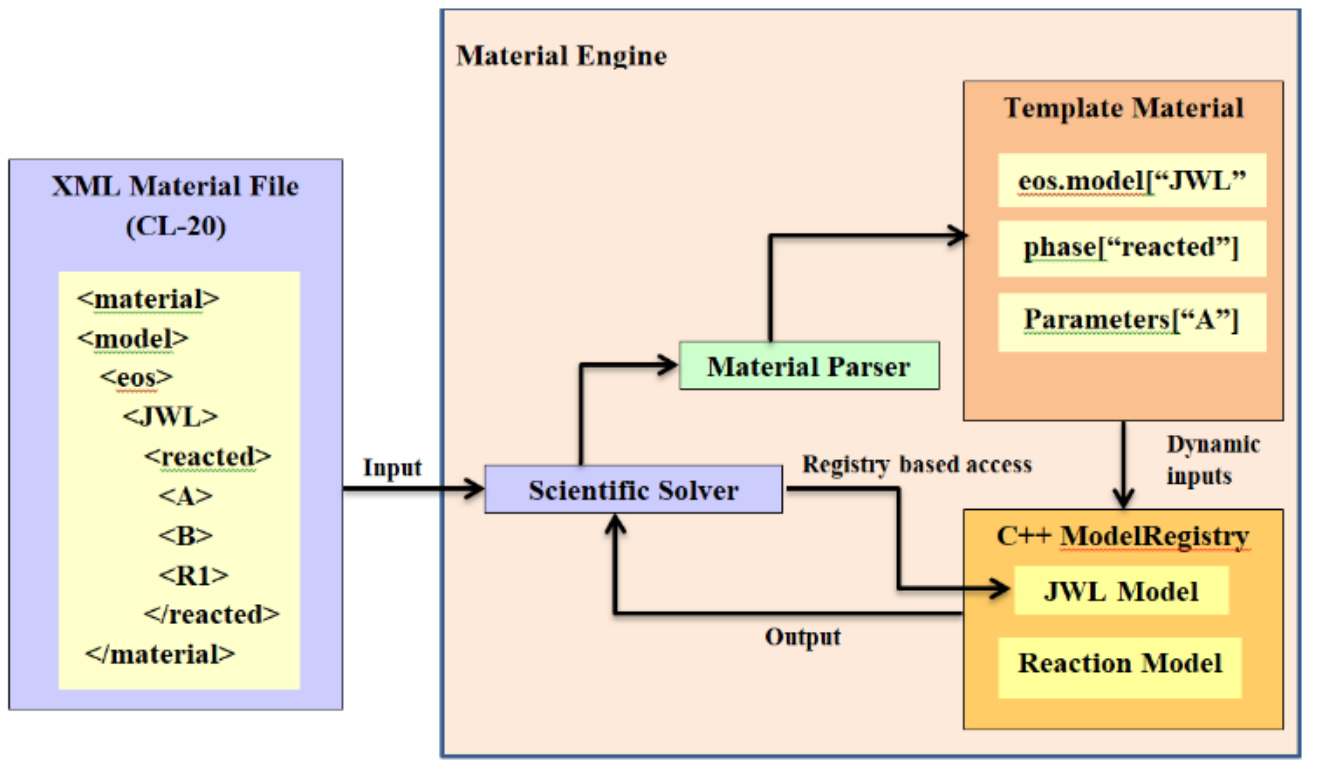

Fig 8: XML-Based Material Parsing and Scientific Solver Integration Workflow

### 5.3 Dynamic Model Identification and Execution

The XML parser dynamically detects and registers models (e.g., JWL, Reaction) at runtime, allowing selection and execution within the simulation loop without recompilation, supporting multiple models seamlessly.

### 5.4 GUI-Driven Material Data Visualization System

The system efficiently manages thermophysical material data, integrating XML, database, AI, and visualization. The GUI, with a Material Browser and Property Viewer, enables centralized access, navigation, and detailed analysis of material properties Fig 9. This section lists all user-modified parameters applied through the GUI. These changes are temporary and do not affect the original database values Fig 10. The Override Management interface allows users to track and manage parameter modifications. It supports sandbox-based experimentation without altering baseline data Fig11.

The Active View displays the current working dataset, combining original and overridden values. It provides a complete and updated representation of parameters with units Fig 12.

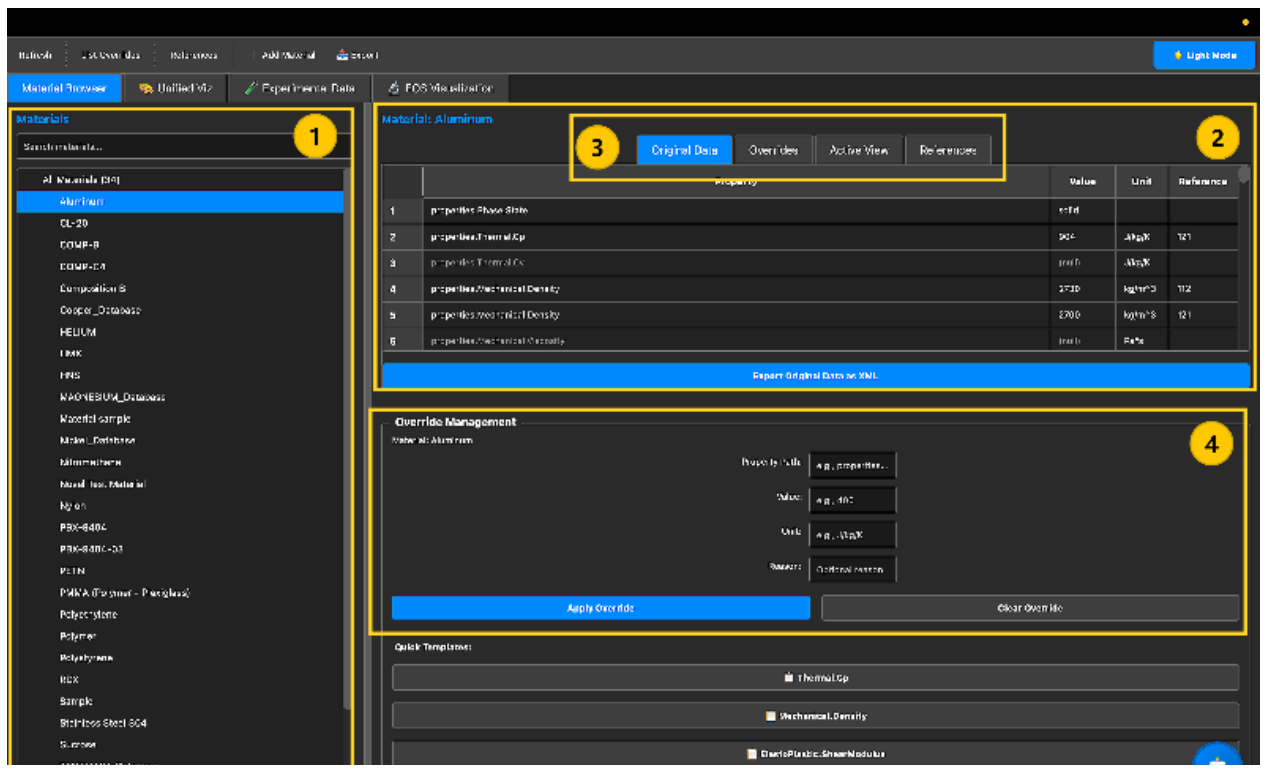

Fig 9: System Main Window with 1) List of Materials; 2) List of properties; 3) Overridden Properties List; 4) Over-ride Management

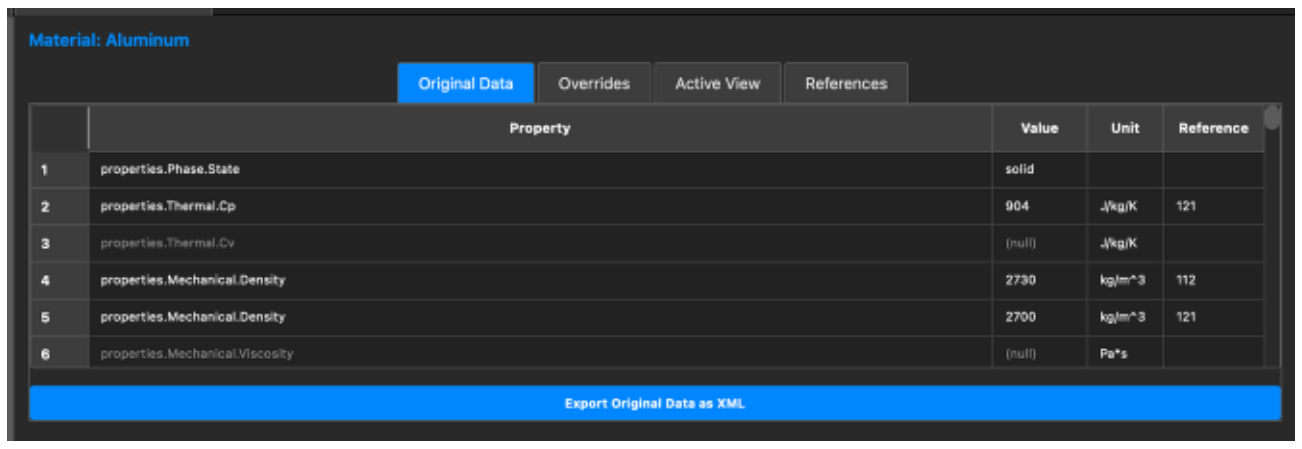

Fig 10: Material Properties, Overrides, Active View and List of References Associated with given materials

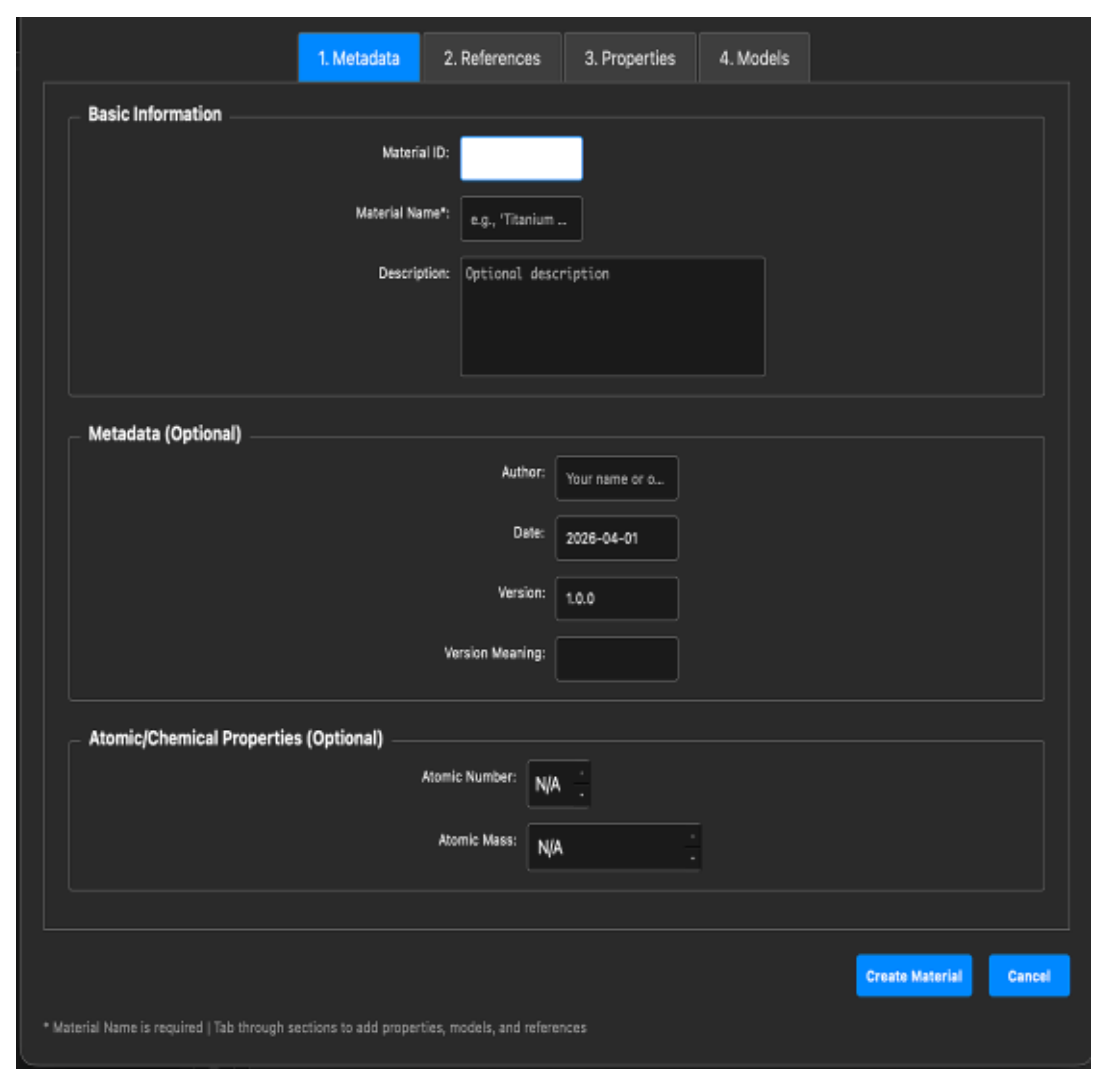

Fig 11: New Material Addition

The AI Chatbot Interface enables users to query the system using natural language. It interprets user input, retrieves relevant data, and presents results in a structured format Fig 12.

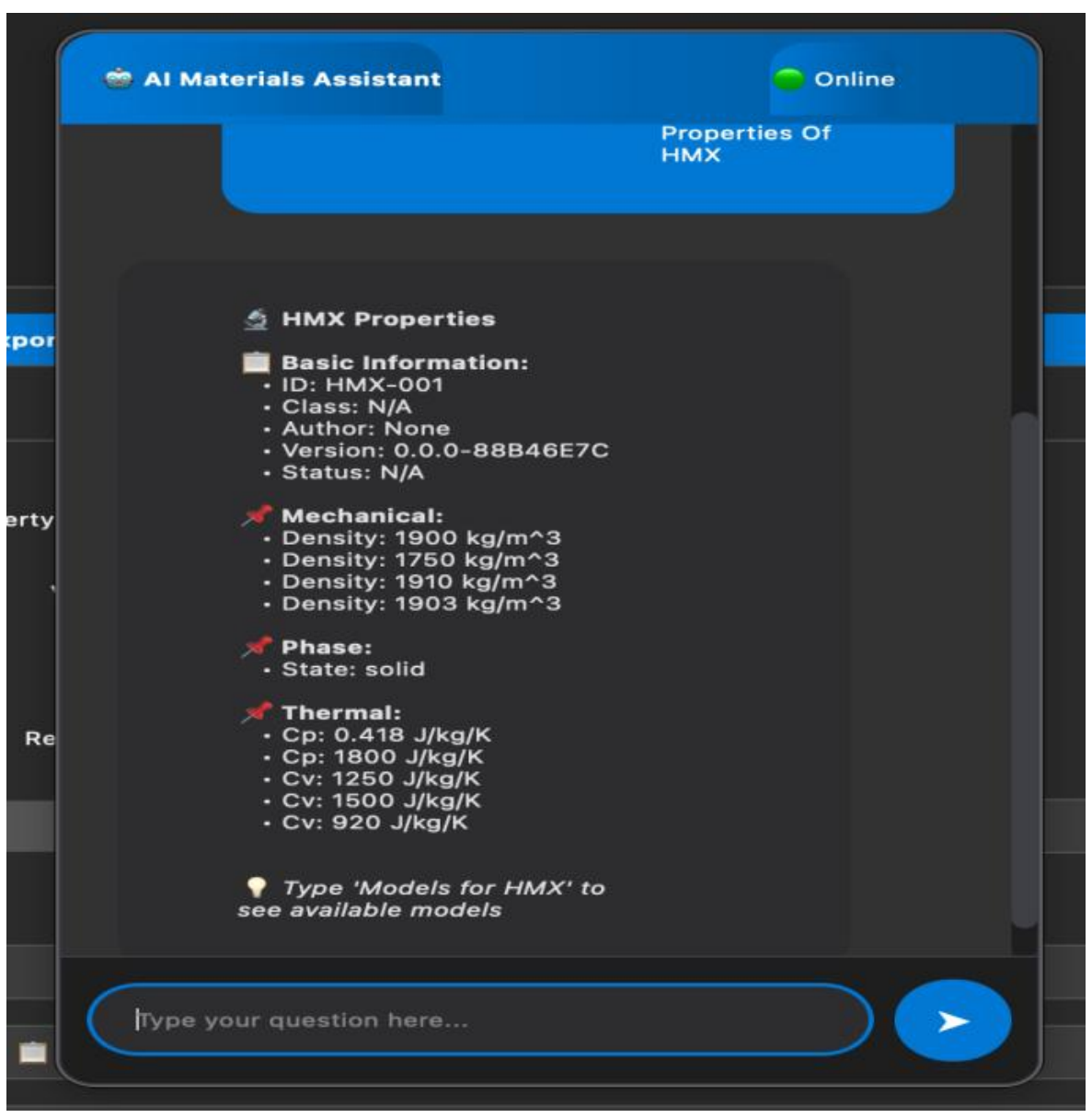

Fig 12: AI Chatbot Interface

### 5.5 Traceable Parameter Flow

The central objective of *Lumina* is to eliminate the “data silos” that traditionally separate quantum-level constants from continuum-scale hydrodynamic physics. While the Mie-Gruneisen Equation of State is used here as a primary case study, the *Lumina* architecture is a generalized framework designed to host any material model—including plasticity, strength, and damage—within a three-level traceable parametric chain.

**Level-1 (Atomistic & Fundamental Inputs):** *Lumina* serves as a repository for the most granular material constants. For the EOS, this includes reference density ($\rho_0$), bulk modulus ($K_0$), and Gruneisen parameters ($\Gamma_0$). However, the framework is built to ingest any "Level-1" data, such as Peierls-Nabarro stresses for plasticity or activation energies for chemical kinetics, sourced directly from Molecular Dynamics (MD) or DFT calculations. (See Table 2).

**Level-2 (Physically-Grounded State Models):** This level acts as the mathematical bridge where atomistic constants are transformed into functional relationships. In this demonstration, the third-order Birch–Murnaghan cold curve serves as the thermodynamic backbone. Because the framework is modular, this level can easily be swapped or expanded to include other "backbone" models, such as Johnson-Cook for strength or P-alpha for porosity, ensuring the physics remains traceable to Level-1.

**Level-3 (Integrated Continuum Response):** The final stage couples the underlying state models with macroscopic physics (e.g., Rankine-Hugoniot relations). Through an automated iterative process, *Lumina* predicts post-shock properties like pressure, temperature, and specific volume. This demonstrates that the platform can translate a microscopic adjustment at Level-1 into a visible shift in the continuum-scale response at Level-3.

### 5.5.1 Key Observations and Framework Scalability

The implementation of the Mie-Gruneisen model for HMX and RDX validates the platform's ability to handle high-fidelity, evolving data:

**High-Resolution Energy Tracking:** By comparing Hugoniot curves to isentropes, *Lumina* provides a clear metric for shock-induced energy loss. This visualization capability is universal; it can be applied to visualize "energy surfaces" for any material model integrated into the platform.

**Comparative Material Analysis:** The platform successfully captures the nuanced differences between HMX and RDX. For example, *Lumina* highlights that while both follow a linear $U_s - U_p$ trend, their thermal signatures diverge significantly due to different lattice vibration modes.

**Sensitivity to Model Complexity:** A critical observation made possible by *Lumina* is the non-linear variation of the Gruneisen parameter in RDX compared to the gradual decrease in HMX. This proves that a "one-size-fits-all" constant parameter is insufficient and justifies *Lumina's* ability to store and execute diverse mathematical representations (e.g., constant vs. polynomial fits) for the same property.

**Internal Consistency & Anisotropy:** Bulk modulus calculations derived from different methods within the framework show high agreement. The platform's sensitivity allows it to detect small low-pressure discrepancies, which are traced back to material anisotropy—demonstrating that *Lumina* doesn't just store data, it understands the physical limits of the models.

**Database-Driven Evolution:** As atomistic models for specific heat ($C_V$) or lattice vibrations ($\Theta$) continue to improve through new MD studies, *Lumina* allows these parameters to be updated at Level-1, instantly propagating those improvements through to the Level-3 continuum predictions.

Table 2: Level-1 Sample Atomistic Input Parameters for HMX and RDX

| Parameters | Symbol | HMX | RDX | Literature |
|---|---|---|---|---|
| **Reference density** | $\rho_0$ (kg/m³) | 1900 | 1806 | Sewell et al, Hooks et al, [14], [15], [16], [17] |
| **Zero-pressure bulk modulus** | $K_0$ (GPa) | 16.5 | 13.0 | Sewell et al, Hooks et al, Zhu et al [14], [15], [16], [17] |
| **Pressure derivative** | $K_0'$ | 8.7 | 9.2 | Sewell et al, Hooks et al, Zhu et al [14], [15], [16], [17] |
| **Reference Grüneisen parameter** | $\Gamma_0$ | 1.1000 | 0.6667 | Sewell et al, Hooks et al, Zhu et al [14], [15], [16], [17] |
| **Grüneisen density coefficient** | $\Gamma_1$ | −0.200 | +2.009 | Sewell et al, Hooks et al, Zhu et al [14], [15], [16], [17] |

## 6. CONCLUSION

This work presents *Lumina*, a unified and scalable informatics framework for managing, analyzing, and retrieving thermophysical material data across multiple scales. The system successfully addresses the challenges of fragmented datasets by integrating hierarchical XML-based data representation, a structured database architecture, and dynamic visualization tools into a single cohesive platform. The modular design enables seamless interaction between data processing, storage, and computational modules, ensuring flexibility and extensibility. The implementation of multiple Equation of State (EOS) models, including $U_s - U_p$, Mie–Grüneisen, and JWL, demonstrates the system's capability to accurately model material behavior under extreme conditions. The integration of a three-layer data management system (Original, Override, and Active views) further supports iterative design, sensitivity analysis, and data-driven experimentation without compromising baseline data integrity. The PyQt6-based GUI enhances usability by providing intuitive access to material data, visualization, and comparative analysis, while the AI-integrated query system simplifies complex data retrieval through natural language interaction. This significantly reduces user effort and eliminates the need for domain-specific query expertise. The results validate that *Lumina* effectively bridges the gap between atomistic data and continuum-scale simulations through a traceable and auditable parameter flow. By combining structured data management, visualization, and AI-driven interaction, the framework establishes a robust foundation for predictive modeling, design optimization, and advanced research in defense and aerospace materials. Overall, *Lumina* represents a significant step toward data-driven material informatics, enabling efficient, accurate, and scalable exploration of material behavior in extreme environments.